# A Leakage-Safe, Cost-Aware Regression Testing Methodology for the Quantum Transpiler

*A Pre-Registered, Cost-Heterogeneous Evaluation of a Transparent Selector Against Simple Baselines for Qiskit*

Furqan Nasir[1,2], Muhammad Arif Shah[1], Iftikhar Alam[1]

[1] City University of Science and Information Technology (CUSIT), Peshawar, Pakistan (PhD Student, Computer Science)

[2] Department of Computer Science, National University of Computer and Emerging Sciences (FAST-NUCES), Islamabad, Pakistan (Lecturer)

Corresponding author: Furqan Nasir (furqannr@gmail.com) · ORCID: Furqan Nasir 0000-0002-5259-3448, Muhammad Arif Shah 0000-0003-0090-3333, Iftikhar Alam 0000-0002-8087-5485

## Abstract

Quantum SDKs such as Qiskit are revised continually, and a single transpiler-pass change can introduce a software regression, so continuous integration (CI) must select and prioritize tests from a large, cost-heterogeneous suite under a fixed budget. We present a leakage-safe, cost-aware regression-test-selection methodology for the Qiskit quantum transpiler, with a pre-registered, budget-binding evaluation that avoids data leakage. The evaluation uses a cost-heterogeneous corpus (116 units, per-unit cost spread 9,945×, $T_{full}$ = 69.17 s) whose budgets were fixed from measured cost before any test-oracle label was read. A transparent risk_score selector does not beat simple test-case-prioritization baselines: mean detection-vs-budget AUC is 0.721 [95% CI 0.44, 0.94] versus 0.874 [0.65, 1.00] for diversity-only (Cliff's $\delta = -0.64$, large), with cost/history/change-stage baselines at 0.840 and random at 0.821. A component ablation shows the composite underperforms its own best signal: diversity and novelty, not severity or cost, are the effective signal. A second, independently pre-registered evaluation (19 mutation-testing events: nine original plus ten new, verified operators) tests the decomposed, diversity-first selector this mechanism implies: it closes the gap to the strongest baseline to a negligible effect size (0.880 vs 0.880, $\delta = -0.08$) while decisively beating the original composite (0.880 vs 0.768, $\delta = +0.57$). Two verified forward-regression events from real Qiskit CI history corroborate it, reported per event, not pooled, under a pre-declared claim-scope rule. We report the negative result as an honest, pre-registered finding with its mechanism. All code, data, and artifacts from this empirical software engineering study are released for reproducibility.

## 1. Introduction

Quantum software development kits (SDKs) such as Qiskit are under rapid, continuous development. The compilers inside them, the transpilers that map an abstract circuit onto a hardware-native gate set and a limited qubit connectivity, are revised very often, and a single change in a layout, routing, translation, optimization, or scheduling pass can introduce a regression. In practice this is a budgeted regression-testing problem. The atomic unit of transpiler testing is a tuple (circuit, backend, transpiler configuration, oracle), and a realistic suite ranges over many circuit families and device topologies. Existing benchmark suites already enumerate on the order of a thousand workloads for a single SDK [1]. Re-executing the full suite with its oracles on every change would quickly exhaust a continuous-integration (CI) budget, since the cost of each test (transpilation plus oracle evaluation) varies widely. So which tests, and in which order, best reveal a regression introduced by an incoming change?

Answering this question well requires a correctness signal that actually notices a regression when one occurs. A companion study shows that the field's de-facto criterion for transpiler correctness, output equivalence modulo global phase and qubit-layout permutation, is systematically incomplete. About 28% of real, merged Qiskit transpiler bug-fixes corrupt layout/permutation contract metadata, introduce non-determinism, or drop the global phase while leaving the compiled output map correct, and this gap replicates on two other compilers [2]. We take that observability gap as motivation and as a design constraint on the oracle used here, rather than re-deriving it. The present paper instead asks how to spend a limited testing budget once an appropriate oracle is already in place.

For conventional software, this budgeted-selection question is already addressed by classical regression test prioritization. Decades of empirical evidence show that simple cost-, history-, and change-aware heuristics are strong baselines, and they frequently stay competitive with much heavier learned models when labelled failure data are scarce [3, 4]. To the best of our knowledge, though, no earlier work has formalized or evaluated cost-aware regression test selection for the quantum transpiler specifically, where the faults are compiler regressions and the oracle must reason about circuit equivalence under a permutation of the qubit layout. Most existing quantum-testing research instead aims at finding bugs and generating tests through metamorphic or differential testing [5, 6, 7, 8], or at benchmarking SDKs [1]. This work is complementary to our problem, but it is distinct from the budgeted selection of existing tests for an incoming change.

This paper's central contribution is a leakage-safe, cost-aware regression-test-selection methodology for the quantum transpiler, and a pre-registered, budget-binding evaluation of a transparent selector within it. Concretely:

1. **A leakage-safe, cost-aware selection methodology (§4).** We separate change events into non-pooled cohorts (forward regression, fix-boundary differential, mutation). We apply a transparent risk_score selector against five baselines and random ordering under budgeted selection, and we enforce six implementation-validity gates and a strict claim-scope rule. Budgets are fixed as declared fractions of measured cost before any label is read, so this is a pre-registered freeze, not a post-hoc tuning exercise.
2. **A cost-heterogeneous corpus that makes the comparison budget-binding, not saturated (§4.3b, §6.1).** A uniform-cost pilot corpus lets every selector tie, because the budget never binds. We design and freeze a cost-heterogeneous corpus (116 units, per-unit cost spread 9,945x, T_full = 69.17 s) whose budgets bind by construction, so selectors are forced to differentiate under real budget pressure. We use “budget-binding” deliberately instead of “powered” in the statistical sense. It describes the corpus design, not a formal power calculation, and the mutation cohort's bootstrap intervals (Table 6) remain wide at n=9.
3. **A budget-binding, honest negative result and its mechanism (§6.3-6.4, §7).** Under the frozen corpus, the proposed risk_score selector does not beat simple baselines (mean AUC 0.721 vs diversity-only's 0.874, Cliff's delta = -0.64, large). A complete component ablation shows why. The multiplicative composite underperforms its own best single signal, and diversity/novelty turns out to be the effective standalone signal: the whole is less than its parts. A post-hoc robustness check (§6.4) shows that this deficit is concentrated in two of the nine events.
4. **A validated fix at scale, independently pre-registered (§6.7).** We evaluate a decomposed selector (diversity first, cost second, severity only as a tie-break) that the ablation mechanism implies, under a fresh, independently frozen 19-event mutation cohort (the original nine operators plus ten new ones, each empirically verified before any label was read). It closes the gap to the strongest baseline to a negligible effect size (Cliff's delta = -0.08 against diversity-only), while decisively beating the original composite selector (delta = +0.57). This is parity with the simplest heuristic at the aggregate level, while still retaining the cost-awareness and severity tie-break that heuristic lacks. We report it alongside the original honest negative, not as a replacement for it.

We answer three research questions. RQ1 (feasibility and rigor) asks whether the leakage-safe pipeline runs end to end, with pipeline-generated labels feeding budgeted selection, while passing all six implementation-validity gates. It also asks whether the cost-heterogeneous corpus design makes the comparison a budget-binding one rather than a saturated tie. RQ2 (selection effectiveness) asks how the transparent risk_score selector compares with five baselines and random ordering on detection-vs-budget AUC, under budgeted CI selection. It asks what effect sizes and confidence intervals result, and what the evidence licenses us to claim given the claim-scope rule. RQ3 (cost/budget behaviour) asks how detection performance varies across the pr/nightly/release budget tiers, and what the frozen cost distribution implies for setting a budget in practice. Observability (§2.2) and the two verified forward-regression events are motivation and supporting evidence for this study, not part of the RQ spine, and any temporal-generalization claim is scoped accordingly (§4.9). Section 2 gives background, and Section 3 surveys related work. Section 4 presents the methodology, Section 5 the implementation, and Section 6 the evaluation. Sections 7–9 discuss findings, threats, and future work, Section 10 concludes, and Section 11 is the artifact statement.

# 2. Background

## 2.1 The Qiskit transpiler

Qiskit compiles an abstract circuit to a target backend through a staged pass manager [9]. The canonical stages are layout, routing, translation, optimization, and scheduling, bundled at optimization levels 0–3. Two properties matter for testing. First, the passes that actually execute depend on the circuit, backend, configuration, and the exact Qiskit revision, so they must be observed rather than inferred from source. Second, transpilation may permute qubits, so a transpiled circuit equals the original only modulo a layout permutation.

## 2.2 Fault types, manifestation channels, and detection metrics

A transpiler regression is classified along two independent axes. The fault-type axis is conventional: correctness faults (a compilation failure or a wrong output), quality faults (a worse two-qubit count or depth), and performance faults (slower compilation). The manifestation-channel axis records how, and whether, a given fault shows up in an observable signal, and it is the more consequential axis for selection. Some channels (output_semantic, circuit_quality, performance, compilation_failure) are visible to a black-box output-equivalence oracle. Others (transpiler_contract_or_metadata, determinism_or_reproducibility, and dropped global phase) are systematically invisible to it by construction, regardless of how the test is selected or ordered. A companion study characterizes this taxonomy in full and estimates the invisible channels at roughly 28% of real Qiskit transpiler bug-fixes [2]. We rely on that taxonomy here only to explain why cohorts are never pooled (§4.2), and why a test's detectability depends on which oracle observed it, not solely on the selector.

Classically, prioritization is measured by APFD and its cost-cognizant variant APFDc, which weights detection by cost and severity under a single shared ordering across multiple faults. That assumption does not hold here, because each change event typically carries one fault, ranked under its own selector-specific ordering, so APFDc is not the natural headline metric. We instead report per-event detection-vs-budget curves, averaged within a cohort and never pooled across cohorts (§4.7).

# 3. Related Work

This paper's true neighbourhood is classical regression test prioritization and selection, not quantum software testing. We position against both and close with an explicit comparison.

## 3.1 Regression test selection and prioritization (RTP)

Rothermel et al. introduced systematic test-case prioritization and the rate-of-fault-detection measure [10]. Elbaum et al. then established APFD, together with a family of empirical studies [11]. The cost-cognizant variant APFDc later incorporated varying test costs and fault severities [12], and it is the conceptual ancestor of the objective in §4.1. In their survey, Yoo and Harman observe that change-, history-, and cost-aware heuristics remain persistently strong baselines [3]. This is exactly the pattern this paper's mutation-cohort result reproduces (§6.3): the transparent risk_score composite does not beat cheapest-first, history-only, or diversity-only. The continuous-integration setting reinforces this pattern. Lightweight selection and prioritization that avoid coverage instrumentation are markedly more cost-effective [13]. A large empirical study of readily available CI and version-control metadata finds that simple, well-known heuristics frequently outperform complex machine-learned models [14]. Information-retrieval-based prioritization reports the same pattern, with simple, well-tuned heuristics staying competitive [15]. This accumulated evidence motivated our choice of a transparent selector over a learned one for this paper's first evaluation, and it now also explains, rather than merely predicts, the result.

Two adaptations are nevertheless required for the quantum transpiler. First, the fault is a compiler regression that spans correctness, quality, and performance, rather than a test failure in the program under test. Second, the oracle must decide circuit equivalence modulo a qubit-layout permutation. We keep the cost-aware objective and the strong-baseline evidence from this literature, but we redefine the fault model, the oracles, and the metrics for this setting (§4).

## 3.2 Learning-based test selection

At industrial scale, predictive (learned) test selection trains gradient-boosted models on large histories of test outcomes, and it can halve testing cost while keeping most of the failure signal [4]. Such methods are data-hungry: they presuppose a large labelled corpus of per-test outcomes, and no such corpus yet exists for quantum-transpiler regressions. This paper's own ledger has nine mutation events and two verified forward events (§6.1, §6.6). A broader literature applies reinforcement learning [16], multi-armed bandits for volatile test pools [17], and other learners surveyed systematically [18] to CI test prioritization, and all of them require substantial historical signal. This reinforces a staged plan: a transparent selector first, with a learned comparison reserved for a later scale-up once enough verified events accumulate, under identical anti-leakage rules (§9).

## 3.3 Quantum software testing

Quantum software testing must contend with the oracle problem [19, 20]. Existing approaches include metamorphic testing (MorphQ [5] and its reproduction MorphQ++ [6]), property-based testing (QuCheck [7]), equivalence-modulo-inputs and differential testing (QEMI [8], QDiff [21], QSPE [22]), concolic testing [23], and formal verification (Giallar [24]). Static analysis and mutation tools round out the landscape (LintQ [25], Bugs4Q [26], Muskit [27], quantum circuit mutants [28]). All of these efforts generate, transform, or verify programs to expose defects, but none of them addresses budgeted selection among an existing test suite for an incoming transpiler change. A companion study surveys this space in full, characterizes how often real transpiler regressions are invisible to output-equivalence oracles, and builds the fault-class-matched oracle family this paper's evaluation relies on [2, 29]. We use its taxonomy and its oracles here (§2.2, §4.4) rather than re-deriving them. Quantum software engineering is by now an established research thread [30], including transpiler optimization itself as an active target [31]. Recent methodological guidance calls for more rigorous empirical reporting in this space [32], and this paper's leakage-safe, gate-audited, claim-scoped methodology (§4.8-4.9) is a direct response to that call.

## 3.4 Positioning

Table 1 positions this paper against the closest threads. Classical RTP/TCP and predictive selection are cost-aware and already select and order existing tests, but neither uses a quantum-aware oracle or a leakage-safe temporal split designed for compiler regressions. The quantum-testing tools surveyed in §3.3 use quantum-aware oracles, but they generate or transform programs to find bugs rather than selecting among an existing budgeted suite. Open QBench [33] benchmarks platform performance, which complements selecting and ordering tests rather than competing with it. Paper 1 [29] builds the quantum-aware oracle family this paper's evaluation depends on, but it does not address budgeted selection at all. To our knowledge, the distinguishing combination here (cost-aware selection and ordering of an existing test suite, under a quantum-aware oracle, with leakage-safe temporal evaluation and non-pooled cohort separation) is individually present in prior work, but it has not previously been combined for the quantum transpiler. (● = yes, ◖ = partial, ○ = no/not applicable.)

**Table 1. Positioning relative to prior work.**

| Approach | Primary goal | QA oracle | Cost-aware | Selects/ orders existing | Leakage-safe temporal | Cohort sep. |
|---|---|---|---|---|---|---|
| Classical RTP/TCP | prioritize tests of a program | ○ | ● | ● | ◐ | ○ |
| Predictive selection | learned selection at scale | ○ | ● | ● | ● | ○ |
| Quantum testing tools (MorphQ etc.) | metamorphic/ differential bug finding | ● | ○ | ○ | ○ | ○ |
| Open QBench | SDK/platform benchmarking | ◐ | ○ | ○ | ○ | ○ |
| Paper 1 (companion) | oracle-observability + oracle family | ● | ○ | ○ | ○ | ○ |
| This work | cost-aware test SELECTION for the transpiler | ● | ● | ● | ● | ● |

# 4. Methodology

## 4.1 Problem formulation

Given a transpiler change and a fixed per-event budget B, select and order tests to maximize early detection of regressions. Conceptually each test is scored by expected_value ≈ P(regression | change, circuit, backend) × impact ÷ expected_cost, and tests are chosen greedily within B. P(·) is the conceptual target. The proposed selector realizes it with a transparent surrogate (§4.5). The primary evaluation unit is the (event_id, test_id) pair.

## 4.2 Change-event model and cohorts

The unit of the dataset is a change event, not a commit. An event is a tuple (baseline_revision, candidate_revision, change_metadata, fault_id), and each one carries a unique fault_id and a fault_type. Every event declares one of three cohorts, and the cohorts are reported separately and never pooled (Figure 1). A forward_regression event pairs the last known-good parent (the baseline) with the regression-inducing commit (the candidate). It is the only cohort that models the real CI question. A fix_boundary_differential event is used in reverse orientation when the introducing commit has not been traced. It pairs a fix commit (the baseline) with its parent (the candidate), which yields a fault-revealing differential. We never describe this differential as a prediction of an incoming change. A mutation event injects a single controlled fault on a base revision, and it is used only to fill gaps in fault-type coverage. Throughout, we prefer historical events.

The mutation operators and the risk_score weight tables (§4.5, Table 2) never share an input, so neither could have shaped the other, regardless of which was authored first. The operators are indexed by pass stage and operator identity (disable_optimization_loop, insert_redundant_cx_pair, and the other seven). The selector's severity and confidence weights, by contrast, are indexed only by circuit family and oracle type, properties of the test corpus and oracle design that are fixed independently of which fault is injected. This matches the fault_type-blind design that the anti-leakage controls require (§4.8). risk_score never takes fault_type, runtime, quality metrics, or an outcome label as input, and an earlier design that keyed severity on fault_type directly was replaced for exactly this reason before evaluation.

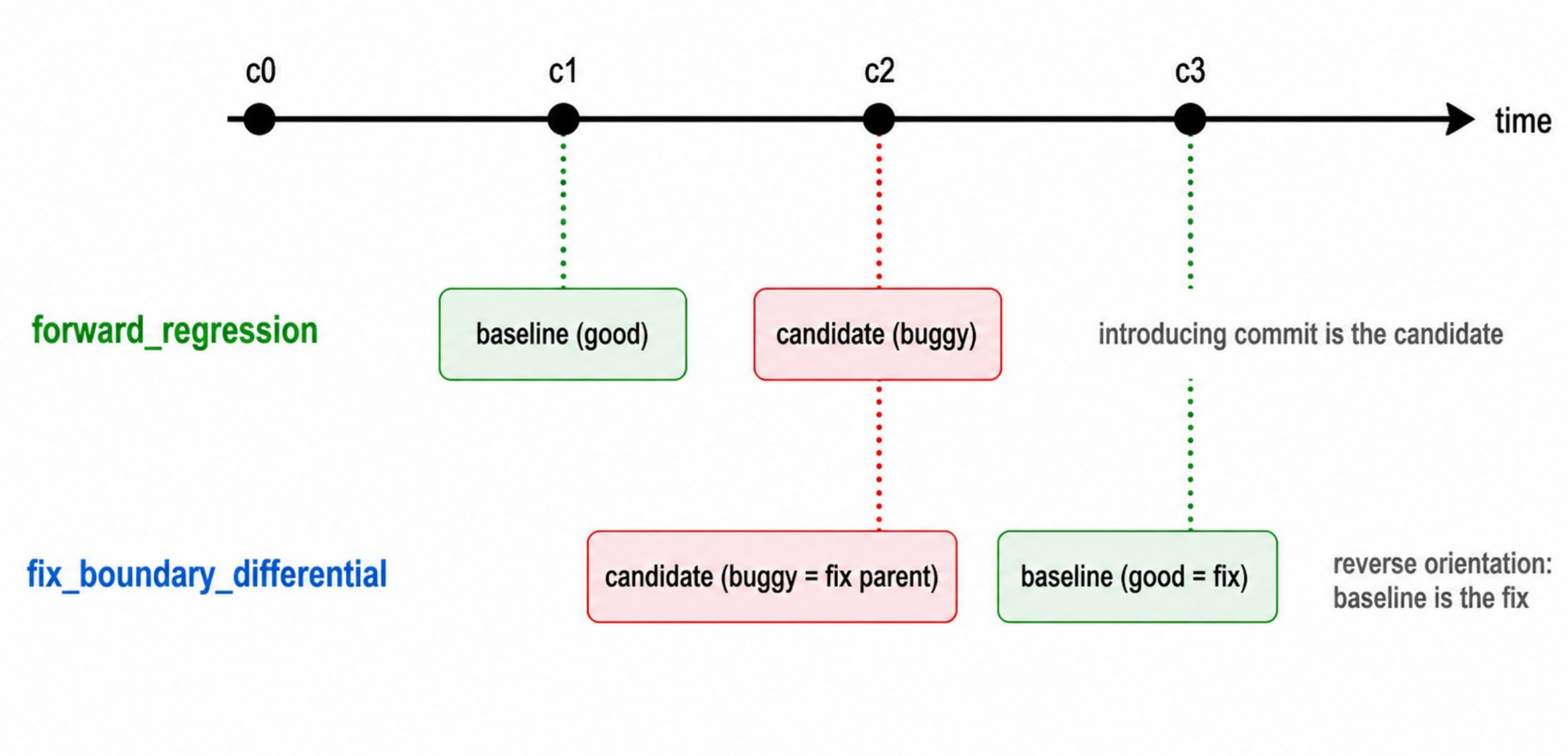

**Figure 1. The three change-event cohorts (forward_regression, fix_boundary_differential, mutation), reported separately and never pooled (§4.2).**

## 4.3 Test units, manifest, and provenance

A test unit is a tuple (circuit, backend, transpiler_configuration, oracle). A static manifest stores its metadata. A separate calibrated profile records the measured baseline cost, the pass stages that actually executed (obtained through instrumentation rather than inferred from filenames), routing observations, and peak memory, keyed by (event_id, baseline_sha, test_id, event_environment_id). Each unit also declares its provenance, one of pre_existing, extracted_from_fix, or synthesized, together with an available_before_candidate flag. These support the two evaluations described in §4.8.

## 4.3b The cost-heterogeneous corpus

A test corpus whose per-unit costs are nearly uniform defeats cost-aware selection by construction. If every candidate test costs about the same, a fixed-fraction budget either admits almost all of them or almost none. Cheapest-first, diversity-only, and a cost-aware composite then all reduce to the same ordering, and the comparison cannot discriminate between them. We therefore design the corpus so that cost varies by orders of magnitude across units. This forces a real trade-off between cheap-but-redundant tests and expensive-but-informative ones once a budget binds.

The corpus enumerates two family groups, cheap_sparse (GHZ, linear-QAOA) and expensive_dense (QFT, random-Clifford), whose transpilation cost diverges sharply with width. It spans four width tiers (small: 5, 8, 12; mid: 18, 27; large: 40, 65; xlarge: 100, dense families only), two backend topologies (line, heavy-hex), and two optimization levels (1, 3), giving 116 test units. The semantic-oracle tier is set by width: an exact unitary check for $n \leq 12$ qubits, a sampled layout-normalized statevector for 13–27 qubits, and a structural fallback beyond that. This way the oracle itself never becomes the cost driver at scale.

Each unit's cost is total_cost_s = baseline_cost_s + oracle_est: the measured transpilation cost plus a per-tier oracle-cost estimate. baseline_cost_s is calibrated as the median of at least five clean CPU

process_time runs per unit, with one warmup run discarded, and measurement is taken outside any profiling instrumentation. (An instrumented single-run measurement was found to inflate cost by roughly 5.5× on a cheap unit during calibration, Appendix F.) Dispersion (min / median / max / IQR) is recorded per unit alongside the point estimate, and a unit is flagged unstable if its IQR exceeds 25% of its median.

Budgets for the pr, nightly, and release CI tiers are declared as fixed fractions of the measured full-suite cost T_full (5%, 50%, and 100% respectively, Table 2). They are set from the cost measurements alone, before any detection label is read for any event. This pre-registered freeze is what makes the subsequent comparison leakage-safe rather than tuned to the outcome. Before the freeze is finalized, each unit is also checked against the frozen budget boundaries for a knife-edge condition, meaning its cost-dispersion range straddles a tier boundary (Appendix F).

## 4.4 Oracles

This section describes the oracles that this evaluation actually exercises: semantic, quality, performance, and an empirical false-positive rate. The full fault-class-matched oracle family for output-invisible faults (a layout/permutation contract differ, contract-level metamorphic relations, and a global-phase tracker) is Paper 1's contribution. We use it here only through its detection labels, without re-deriving it.

**Semantic.** The semantic oracle is tiered by applicability and width. For a small circuit (at most 12 qubits) that is unitary-pure and ancilla-free, it checks exact unitary equivalence modulo global phase after first normalizing the layout. For every other circuit it falls back to structural validity, meaning adherence to the basis and to the coupling map. A full $2^n$ operator is therefore never materialized for large circuits. This mirrors dedicated quantum equivalence-checking practice, which compares the maps of the pre- and post-compilation circuits and exploits reversibility and simulation to stay tractable [34].

**Quality.** A Pareto rule over pre-declared thresholds governs quality regressions, with Δ two-qubit count and Δ depth reported separately. A sensitivity grid is reported, and thresholds are fixed before held-out evaluation.

**Performance.** A two-stage protocol, a median-slowdown screen followed by a robust effect size, runs exploratory on a laptop, timed with CPU process time.

**Empirical false-positive rate.** A warning is a false positive only if not upheld by independent confirmation, never merely because an oracle fired.

### 4.5 The proposed transparent selector (risk_score)

The selector introduces no ML and outputs an interpretable score, not a calibrated probability (Figure 2):

$$\text{risk_score} = (\varepsilon + \text{change_stage_match}) \times \text{circuit_backend_sensitivity} \times \text{impact_prior} \times \text{oracle_confidence} \div \text{expected_cost} \times \text{novelty_multiplier}$$

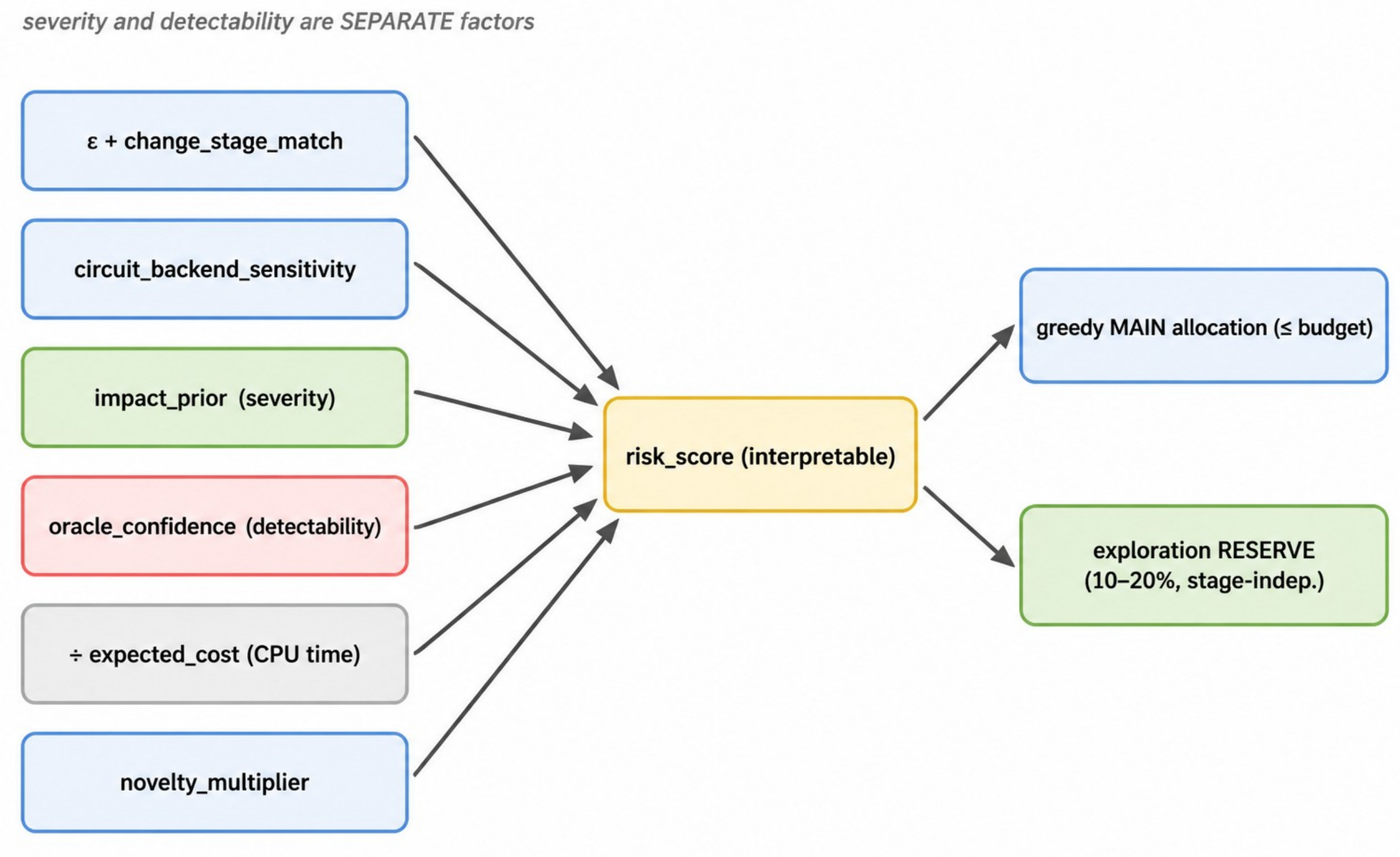

**Figure 2. The risk_score selector: a transparent, multiplicative composite of change-stage match, circuit/backend sensitivity, impact prior, oracle confidence, expected cost, and a novelty multiplier (§4.5).**

Fault severity (impact_prior) and detection confidence (oracle_confidence) enter as separate multiplicands, so a high-impact test that relies on a weaker oracle is not penalized twice. The ε floor, together with a stage-independent exploration reserve of 10–20%, keeps an incomplete change–stage map from driving useful tests to zero. Unknown or framework-wide changes are routed into that reserve, and when no prior qualifying events exist, a neutral cold-start fallback applies. The selector sees only pre-execution information. The candidate's fault_type, runtime, quality, and labels are never inputs.

**Table 2. Frozen risk_score configuration (pre-declared before evaluation).**

| Parameter | Value / rule |
|---|---|
| epsilon (change-stage floor) | 0.05 |
| exploration reserve | 15% of budget (admissible range 10–20%) |
| novelty decay (per repeated family×backend) | 0.5^count |
| circuit criticality (impact_prior) | GHZ 1.0, QFT 1.5, linear-QAOA 1.3, random-Clifford 1.2 |
| oracle_confidence | unitary 1.0, sampled-SV 0.8, property 0.5, structural 0.3 |
| oracle cost estimate (s) | unitary 0.05, sampled-SV 0.20, property 0.05, structural 0.01 |
| circuit_backend_sensitivity | (1+pressure) × (1+two-qubit density) × (1 + n/30) |
| impact_prior | criticality × (1+pressure) × (1 + prior-failure-rate) |
| cold-start | prior-failure-rate = 0 ⇒ impact factor 1.0 |
| tie-breaking / determinism | ties broken by test_id, seed-deterministic ranking |

### 4.6 Baselines and budgeted selection

Five baselines observe the same budget: random, cheapest-first, diversity-only, history-only, and the change-stage heuristic [3]. Selection reserves 10–20% of the budget for stage-independent exploration, then fills the main allocation greedily by score. It ensures at least one representative per circuit family where the budget permits, never exceeds the budget, and is seed-deterministic.

The baselines behave as follows. Random applies a seeded uniform shuffle of the candidate units. Cheapest-first orders them by ascending expected cost. Diversity-only greedily covers (circuit family × backend) combinations in ascending order of cost, and then adds the cheapest remaining units. History-only ranks units by descending prior failure rate, computed only from events strictly before the event date, breaks ties by cost, and reduces to cheapest-first under a cold start. Change-stage prefers units whose declared pass stages include the modified stage, again breaking ties by cost.

### 4.7 Metrics

Per event and averaged, cohorts/sources separated: Recall@budget, normalized TTFR, detection rate at {5,10,20,40}% budget, area under the detection-vs-budget curve, selection overhead, diversity coverage, and the empirical oracle FP rate. APFDc is reported only for the atypical multiple-fault-per-ordering case.

### 4.8 Anti-leakage controls

Four controls enforce leakage safety. The first is a group-level temporal split keyed by split_group_id = base_revision + mutation_family. Any group that straddles the cutoff goes entirely to the test side. The second is provenance-driven. A Primary CI evaluation restricts the candidates to tests that were available before the candidate, while post-fix targeted triggers are confined to a Secondary retrospective evaluation. The third freezes the change–stage map before held-out evaluation, and any later refinement is performed on training events only and is versioned. The fourth calibrates cost with CPU process time (time.process_time) and keeps wall-clock time only as a secondary reference.

### 4.9 Validity gates and claim scope

Six gates check implementation correctness before any result is reported: budget compliance, reproducibility, absence of leakage, label reproducibility from the write-once artifacts, metric correctness against hand-computed cases, and oracle coverage. A claim-scope rule then forbids any headline temporal-generalization claim while fewer than three forward_regression events are verified in the held-out cohort. In that situation, results are reported per event, as pilot evidence. Whether the selector actually beats the baselines is treated as the empirical outcome, never as a precondition for reporting.

## 5. Implementation

The methodology is realized in a Python prototype, cart, organized by phase (manifest, events, oracles, labels, selectors, metrics, gates) with a CLI (manifest, events, ground-truth, select, evaluate, analyze, historical-run), shown in Figure 3.

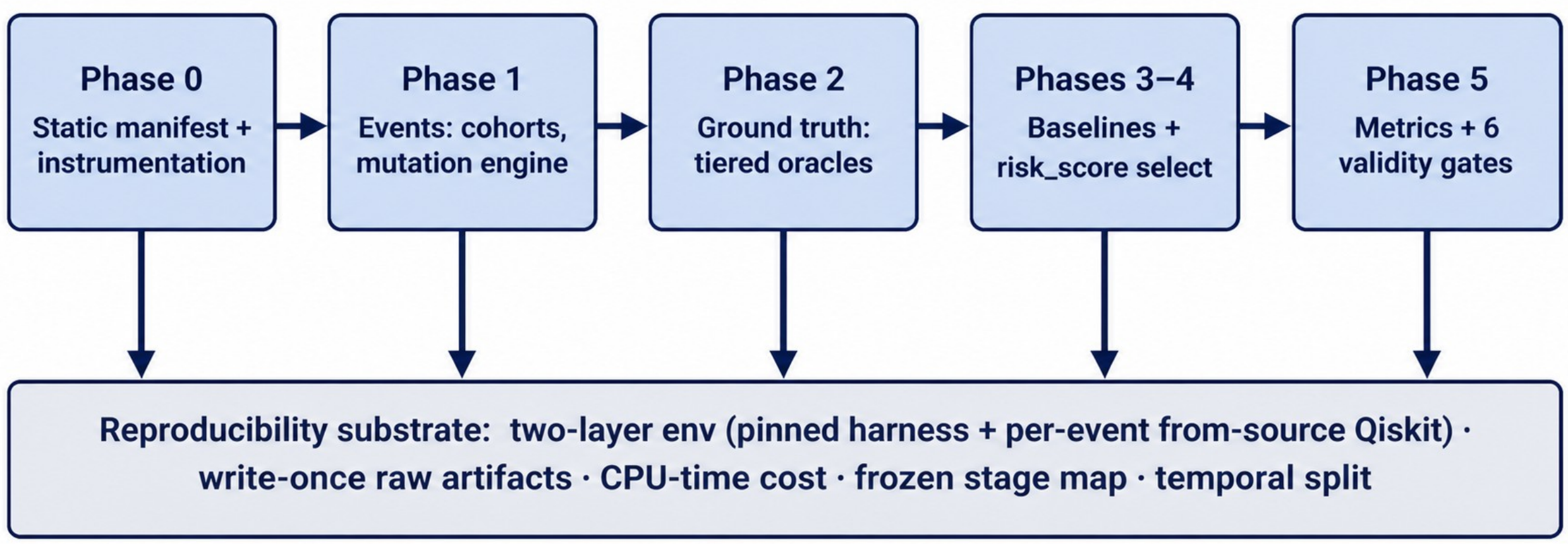

**Figure 3. The cart prototype's pipeline: manifest, events, oracles, labels, selectors, metrics, and gates (§5).**

**Two-layer environment.** Reproducing historical regressions means building several different Qiskit revisions, which is incompatible with a single pinned Qiskit. We therefore separate a fixed harness layer (Python 3.11, Benchpress, and the oracle and measurement stack, all lockfile-pinned) from a Qiskit-under-test layer that is built from source for each event. Instrumentation is anchored on Qiskit 2.4.2 within the 2.x window. Building from source requires a Rust toolchain.

**Per-event from-source runner.** For historical events, a runner builds the baseline and candidate revisions into isolated virtual environments. In each environment a worker runs as a subprocess, transpiles the test units with its own Qiskit, and serializes the outputs (as QPY) together with the CPU timings. The harness then loads both sets of results back and applies the shared oracle pipeline. In this way the system under test stays isolated, while a single oracle implementation is reused for both revisions.

**Reproducibility artifacts.** Raw oracle evidence is written once and never overwritten, and the derived labels are regenerated from it so that they reproduce exactly, including the timing-dependent fields. The per-event environment recipes, the seeds, the pre-declared thresholds, the frozen stage map, and the frozen per-run configurations are all committed. The prototype also ships an automated test suite, with the six validity gates wired directly into evaluate.

**What is implemented versus exercised in this pilot.** To avoid over-stating coverage we separate three states.

- **Implemented and exercised in the primary mutation evaluation:** the static manifest and instrumented profiling, including the cost-heterogeneous corpus generator (§4.3b); the mutation engine (nine semantics-preserving quality operators); the tiered semantic oracle (exact ≤ 12 qubits, sampled statevector 13–27, structural above); the quality Pareto oracle; the five baselines and the proposed risk_score selector; budgeted selection under the frozen pr/nightly/release tiers; the metric suite; the six validity gates; the ablation/sensitivity analysis (`analyze`); and the held-out evaluation on the nine-event × 116-unit cost-heterogeneous mutation cohort.

- **Implemented and exercised in the secondary historical validation:** the per-event from-source runner; the fault-class-matched oracle family (the layout/permutation contract differ, the metamorphic MR-1 [permutation-consistency] oracle, and the global-phase tracker), and the source verification of H1 (ElidePermutations #14603), H2 (final_layout composition #14919), and the CommutativeCancellation global-phase fix #14956, on each of which the black-box output oracle is blind while the matched oracle detects the fault; the targeted-trigger units and the Secondary retrospective evaluation; and the exploratory performance Stage-1 screen. The determinism (#14730) event remains unreproduced at this scale. The small-overhead performance event (#14120, H4) is confirmed as a forward-regression (§6.6).
- **Deferred:** a noisy-target determinism oracle; a broader sampled-statevector campaign at scale (13–22 qubits); a controlled-hardware, three-run-median performance protocol; and any learned selector.

# 6. Evaluation

We evaluate the three research questions on a single machine, anchored on Qiskit 2.4.2, using the frozen cost-heterogeneous corpus (§4.3b). We state the freeze order explicitly because it is the basis for every leakage-safety claim in this section. The corpus was enumerated, costs were measured, and budgets were set from those costs alone, all before any detection label was read for any event (configs/corpus_costhet.yaml, version 1, frozen: true, dated 2026-08-07).

## 6.1 Corpus and setup

The corpus enumerates 116 test units: two family groups (cheap_sparse = GHZ, linear-QAOA; expensive_dense = QFT, random-Clifford) across four width tiers (small: 5, 8, 12; mid: 18, 27; large: 40, 65; xlarge: 100, dense families only), two backend topologies (line, heavy-hex), and two optimization levels (1, 3). Table 3 summarizes this. Hardware: Intel Xeon @2.20GHz, 1 logical CPU, Python 3.12.13, Qiskit 2.4.2, RAYON_NUM_THREADS=1 (§4.3b, Appendix F).

**Table 3. Cost-heterogeneous corpus composition (116 units).**

| Dimension | Values | Count |
|---|---|---|
| Circuit family group | cheap_sparse (GHZ, linear-QAOA); expensive_dense (QFT, random-Clifford) | 4 families |
| Width tier (qubits) | small: 5,8,12; mid: 18,27; large: 40,65; xlarge: 100 (dense only) | 8 widths |
| Backend topology | line, heavy-hex | 2 |
| Optimization level | 1, 3 | 2 |
| Semantic oracle tier | exact-unitary (n<=12), sampled statevector (13-27), structural (>=28) | 3 tiers |
| Total units | | 116 |

Per-unit cost (total_cost_s = baseline_cost_s + oracle_est, §4.3b) spans 0.0008 s to 8.292 s, a 9,945x spread, with median 0.0084 s. The full-suite cost is T_full = 69.17 s over 1,044 calibration profile records (116 units x 9 mutation events). Budgets for the pr, nightly, and release CI tiers are declared as fixed fractions of T_full (5%, 50%, and 100% respectively), set from these cost measurements alone, before any detection label is read (Table 4). The corpus binds by construction. The pr budget (3.46 s) admits 91 of 116 units and excludes 25 of the 29 costliest (the top cost quartile), and the nightly budget (34.59 s) admits 110 of 116 and excludes 6 of the 29 costliest. This is the corpus-design fix for the earlier pilot's saturation problem, where a uniform-cost corpus let every selector tie because the budget never bound. Here the budget applies real pressure on the expensive tail. Appendix F reports cost-measurement stability (0 unstable units, 0 knife-edge units at the frozen boundaries).

**Table 4. Cost distribution and frozen CI-tier budgets (set from costs only, before any label was read).**

| Statistic | Value | Note |
|---|---|---|
| minimum per-unit cost | 0.0008 s | |
| median per-unit cost | 0.0084 s | |
| maximum per-unit cost | 8.292 s | |
| cost spread (max/min) | 9,945x | |
| T_full (full-suite cost) | 69.17 s | 116 units, single pass |
| pr budget | 3.46 s | 5% of T_full, admits 91/116, excludes 25/29 costliest |
| nightly budget | 34.59 s | 50% of T_full, admits 110/116, excludes 6/29 costliest |
| release budget | uncapped | 100% of T_full |

## 6.2 RQ1: Feasibility and rigor

All six implementation-validity gates are independently re-verified against the frozen corpus (Table 5). Five are unconditional passes: budget compliance, reproducibility (identical seed implies identical ranking), leakage absence (a fabricated current-event outcome cannot change a ranking), metric correctness (against hand-computed cases), and oracle coverage. The sixth, label reproducibility, checks that every derived label reproduces exactly from its write-once raw artifact. A second Kaggle calibration pass (KAGGLE_CALIBRATION.md) retained these raw per-unit artifacts, and 1,040 of 1,044 mutation-cohort labels (99.6%) reproduce exactly from an independently re-measured raw artifact. The four exceptions are all instances of the same two mutation types already isolated in §6.4 as the selector's weak point (insert_redundant_cx_pair, insert_redundant_x_pair), applied to circuits costing 1-3 ms to transpile. The mutation cohort's performance-regression flag is a single-pass 20% slowdown threshold with no bootstrap confirmation (§4.4), and 20% of one to three milliseconds sits inside ordinary process-time jitter even under the single-thread pin (§4.3b). No mismatch was observed on any of the other 1,040 records across the full cost range, so this is a measurement-noise boundary specific to the cheapest units in the corpus, not a labeling defect. The frozen labels and the comparisons reported in §6.3-6.6 are unaffected by this re-verification pass, because they were fixed beforehand, per the pre-registration discipline in §4.8.

**Table 5. Implementation-validity gates, independently re-verified 2026-08-08 against the frozen corpus and its raw write-once artifacts.**

| Gate | Result |
|---|---|
| Budget compliance | PASS |
| Reproducibility (same seed => same ranking) | PASS |
| Leakage absence (current-event outcome cannot change ranking) | PASS |
| Metric correctness vs hand-computed cases | PASS |
| Oracle coverage | PASS |
| Label reproducibility from raw artifacts | 1,040/1,044 exact (99.6%). 4 |

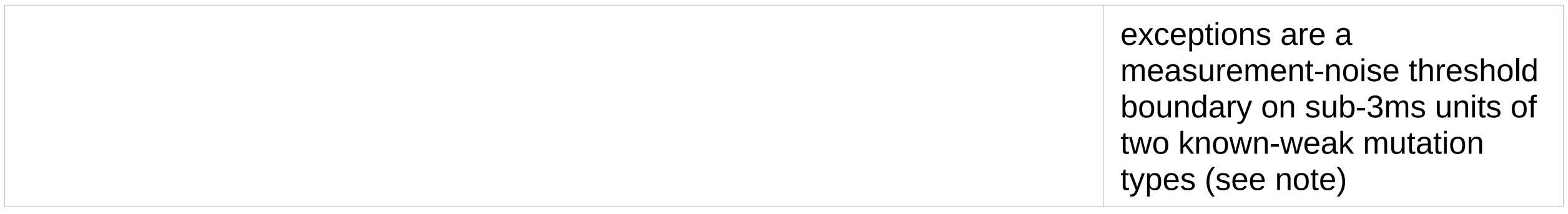

| | |
|---|---|
| | exceptions are a measurement-noise threshold boundary on sub-3ms units of two known-weak mutation types (see note) |

### 6.3 RQ2: Selection effectiveness — the headline

On the mutation cohort (nine events, 116 units each, frozen budgets), the proposed risk_score selector does not beat the simple baselines. Mean detection-vs-budget AUC is 0.7214 [95% CI 0.4389, 0.9444] for the proposed selector, against 0.8742 [0.6520, 1.0000] for diversity-only (the best baseline), 0.8395 [0.6197, 0.9647] for cheapest-first, history-only, and change-stage, and 0.8211 [0.7962, 0.8459] for random ordering over 20 seeds. The three baselines are identical on this corpus because history-only cold-starts to cheapest-first with no prior events (§4.6), and change-stage's stage signal is inert here (§6.4), so all three collapse to the same cost ordering. The effect size against diversity-only is large and negative (Cliff's delta = -0.642, A12 = 0.179). Against cheapest-first, history-only, and change-stage it is small and positive (delta = +0.247, A12 = 0.624). Against random it is small and negative (delta = -0.148, A12 = 0.426). This is an honest, pre-registered negative result for the proposed selector, and we report it as the headline finding rather than spinning it as a win.

**Table 6. Selector comparison on the mutation cohort (9 events, 116 units each). Mean AUC with 95% bootstrap CI. Random is the median over 20 seeds.**

| Selector | mean AUC | 95% CI | recall@20% |
|---|---|---|---|
| diversity_only | 0.8742 | [0.6520, 1.0000] | 0.8889 |
| cheapest_first | 0.8395 | [0.6197, 0.9647] | 0.8889 |
| history_only | 0.8395 | [0.6197, 0.9647] | 0.8889 |
| change_stage | 0.8395 | [0.6197, 0.9647] | 0.8889 |
| random (median, 20 seeds) | 0.8211 | [0.7962, 0.8459] | 0.7500 |
| proposed (risk_score) | 0.7214 | [0.4389, 0.9444] | 0.6667 |

The per-event view explains the mean. On six of the nine mutations, proposed reaches its detecting unit earlier than cheapest-first, history-only, and change-stage (per-event AUC 0.98-0.997), though diversity-only matches or beats it on five of those six (Table 7). It ties on the one mutation the black-box oracle never detects (drop_vf2_post_layout, every selector scores 0.000). It loses catastrophically on exactly two mutations, insert_redundant_cx_pair (0.525) and insert_redundant_x_pair (0.000), where every deterministic baseline scores 0.976-1.000. This 6-win/1-tie/2-loss pattern (Table 7, and the win/tie/loss counts in Table 8) is the same qualitative pattern the earlier pilot corpus showed on the identical two mutation types. That is why the ablation (§6.4) targets the severity term directly: on these two events, a high-impact, high-criticality unit is promoted ahead of the cheap unit that actually detects the fault.

**Table 7. Per-event detection-vs-budget AUC by selector (9 mutation events, 1.0 = detected at near-zero cost, 0.0 = never detected).**

| Mutation event | cheap. | divers. | hist. | ch-stage | proposed | random |
|---|---|---|---|---|---|---|
| disable_optimization_loop | 0.911 | 1.000 | 0.911 | 0.911 | 0.997 | 0.996 |
| drop_vf2_post_layout | 0.000 | 0.000 | 0.000 | 0.000 | 0.000 | 0.000 |
| downgrade_routing | 0.936 | 1.000 | 0.936 | 0.936 | 0.997 | 1.000 |
| insert_redundant_cx_pair | 1.000 | 1.000 | 1.000 | 1.000 | 0.525 | 0.995 |
| drop_optimization_stage | 0.908 | 1.000 | 0.908 | 0.908 | 0.996 | 0.995 |
| force_opt_level_1 | 0.868 | 0.868 | 0.868 | 0.868 | 0.984 | 0.994 |
| force_opt_level_0 | 0.979 | 1.000 | 0.979 | 0.979 | 0.997 | 1.000 |
| downgrade_layout_trivial | 0.977 | 1.000 | 0.977 | 0.977 | 0.997 | 1.000 |
| insert_redundant_x_pair | 0.976 | 1.000 | 0.976 | 0.976 | 0.000 | 0.769 |

Table 8 reports two measurements side by side that are related but not computed the same way, and the distinction matters for reading the table correctly. The Cliff's delta and A12 columns compare the proposed selector's per-event AUC against each baseline's AUC across all cross-event pairs (unpaired). The win/tie/loss columns instead compare the same nine events matched pair-by-pair on normalized TTFR (lower is better), a different metric entirely. The two tell a consistent story here: the same two mutation types drive both the AUC-based effect size and the TTFR-based losses. But the magnitudes are not directly comparable. By the matched TTFR comparison, proposed wins more individual events than it loses against every baseline except diversity-only. Yet its unpaired Cliff's delta against cheapest-first, history-only, and change-stage is only "small" (+0.247), because the all-pairs computation weighs the six narrow per-event AUC wins equally with the two large losses, rather than letting the win count dominate.

**Table 8. Effect sizes (Cliff's delta, Vargha-Delaney A12, proposed vs each baseline on per-event AUC, unpaired all-pairs) and win/tie/loss (proposed vs each baseline, by per-event normalized TTFR, matched by event, 9 events).**

| Comparison (proposed vs) | Cliff's delta | magnitude | A12 | win | tie | loss |
|---|---|---|---|---|---|---|
| diversity_only | -0.642 | large | 0.179 | 1 | 1 | 7 |
| cheapest_first | +0.247 | small | 0.624 | 6 | 1 | 2 |
| history_only | +0.247 | small | 0.624 | 6 | 1 | 2 |
| change_stage | +0.247 | small | 0.624 | 6 | 1 | 2 |
| random (median) | -0.148 | small | 0.426 | 2 | 1 | 6 |

*Effect sizes: per-event AUC, unpaired, all 9x9 cross-event pairs. Win/tie/loss: per-event normalized TTFR, matched by event.*

## 6.4 Ablation and sensitivity of risk_score

To locate the source of the deficit, we ablate each risk_score component over the cached labels, with no re-transpilation (Table 9, results/analyze-costhet-frozen.json). Removing the severity term impact_prior raises the mean AUC from 0.7214 to 0.7248, the single largest improvement from any one-component removal, confirming the mechanism identified on the earlier pilot corpus. This is still far below diversity-only (0.8742) or cheapest-first (0.8395), though, so the severity term alone does not explain the full deficit. Removing novelty lowers the AUC to 0.6946, and removing the cost divisor lowers it to 0.6968. This rules out a cost-division artifact as the cause: cost helps, it does not hurt. change_stage_match and oracle_confidence are inert on this corpus (0.7214 and 0.7213), because most units in the affected events share the modified stage and the same oracle tier. A sensitivity sweep over the epsilon floor ({0.01, 0.05, 0.10}) and the exploration reserve ({10, 15, 20}%) leaves the mean AUC unchanged at 0.7214, so the result is robust rather than a tuning artifact.

**Table 9. Ablation of risk_score components (mutation cohort, 9 events, 116 units, reuses cached labels, no transpilation).**

| risk_score variant | mean AUC | mean nTTFR | recall@20% |
|---|---|---|---|

| full | 0.7214 | 0.2786 | 0.6667 |
|---|---|---|---|
| - novelty | 0.6946 | 0.3054 | 0.6667 |
| - change_stage | 0.7214 | 0.2786 | 0.6667 |
| - oracle_confidence | 0.7213 | 0.2787 | 0.6667 |
| - impact_prior | 0.7248 | 0.2752 | 0.5556 |
| - cost (ranking divisor only) | 0.6968 | 0.3032 | 0.6667 |

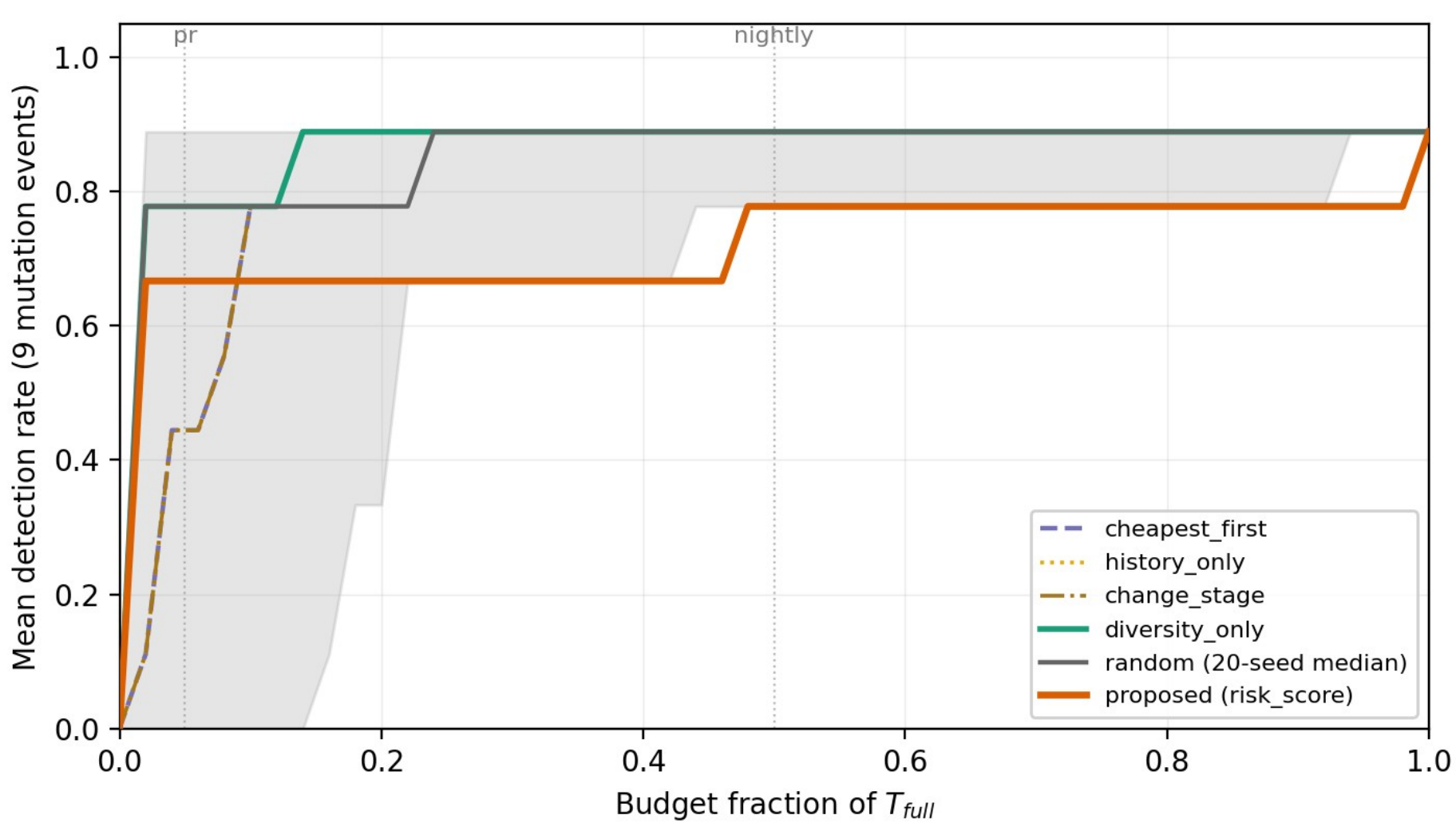

**Figure 4. Mean detection-vs-budget curves by selector on the mutation cohort (9 events). The shaded band is random ordering's 20-seed min-max range (§6.4).**

Figure 4 plots the mean detection-vs-budget curve. Diversity-only and random reach the 8/9 = 0.889 ceiling first. (The ninth mutation, drop_vf2_post_layout, is undetected by every selector because the black-box quality oracle does not register it on this corpus.) Cheapest-first, history-only, and change-stage climb more slowly to the same ceiling. The proposed selector plateaus at 0.667 through the pr tier and 0.778 through the nightly tier, only reaching the shared ceiling at the uncapped release tier. No selector separates from any other beyond the 0.889 ceiling, because that ceiling is an oracle limitation, not a selection-budget limitation. The shaded band shows the 20-seed min-max range for random ordering.

Taken together, the ablation points to a consistent mechanism: the multiplicative risk_score composite underperforms its own best single signal. Removing any one component never lifts the AUC above 0.7248, while diversity-only alone reaches 0.8742. The whole is less than its parts, and diversity/novelty, not severity, not cost, not change-stage matching, is the effective standalone signal on this corpus (§7 returns to why).

As a post-hoc robustness check (not part of the pre-registered comparison, and computed only by re-aggregating the already-frozen per-event numbers with the same metric functions: scripts/robustness_leave2out.py, results/robustness_leave2out.json), we recompute the same selector comparison after excluding exactly the two mutation types that drive the deficit (insert_redundant_cx_pair, insert_redundant_x_pair; n=7 remaining events). On those seven events, proposed's mean AUC rises to 0.852 [95% CI 0.57, 1.00]. This now exceeds cheapest-first, history-only, and change-stage's 0.797 [0.52, 0.95] by a large positive effect size (Cliff's delta = +0.735), and it is statistically indistinguishable from random (delta = -0.122, negligible). Against diversity-only, however, the effect size stays large and negative (delta = -0.490). This is because diversity-only scores exactly 1.000 on five of the seven remaining events, while proposed scores just under it (0.984-0.997). The all-pairs Cliff's delta weighs each of those narrow losses the same as it weighed the two catastrophic ones, so a higher mean AUC for proposed does not translate into a favorable delta against a baseline that rarely loses even a fractional comparison. This sharpens rather than overturns the mechanism above. Essentially the entire mean-AUC deficit against every baseline except diversity-only traces to the two known-weak mutation types, while the deficit against diversity-only specifically is more structural. We report this as a diagnostic, exploratory finding. The pre-registered, frozen nine-event comparison in §6.3 remains the paper's primary and only claimed result.

## 6.5 RQ3: Cost/budget behaviour across tiers

Reading the same detection-vs-budget data (Table 10) at the three declared CI-tier fractions shows where the proposed selector's deficit actually bites. At the pr tier (5% of T_full, the tightest and most CI-realistic budget), proposed already beats cheapest-first, history-only, and change-stage (0.667 vs 0.444), but it trails diversity-only and random (0.778). At the nightly tier (50%), proposed is the worst of the six (0.778, versus 0.889 for every other selector), because its two catastrophic losses (§6.3) are not yet resolved at that budget fraction. At the release tier (100%, uncapped), every selector converges to the same 0.889 ceiling. This confirms that ordering cannot matter once the budget is large enough to admit every unit that the oracle can ever detect: the ceiling is oracle-limited, not selector-limited. The empirical oracle false-positive rate over the full corpus is 0.0059.

**Table 10. Mean detection rate at the three frozen CI-tier budgets (9 mutation events).**

| CI tier (budget) | cheap./hist./ ch-stage | diversity_only | random (med.) | proposed | ceiling reached? |
|---|---|---|---|---|---|
| pr (5% of T_full) | 0.444 | 0.778 | 0.778 | 0.667 | no |
| nightly (50% of T_full) | 0.889 | 0.889 | 0.889 | 0.778 | yes (all but proposed) |
| release (100%, uncapped) | 0.889 | 0.889 | 0.889 | 0.889 | yes (all selectors) |

## 6.6 Forward cohort (per event, n=2)

Two forward_regression events are verified from source, under the claim-scope rule (§4.9). Fewer than three such events means we cannot make a pooled comparative-effectiveness or temporal-generalization claim, so both are reported per event, with no cohort-level aggregate statistic, no pooled confidence interval, and no cross-event Cliff's delta. The within-event confirmation statistics reported for H4 below (a multi-run bootstrap CI and Cliff's delta comparing its own buggy and fixed builds) are a single event's own evidence, not a pooled cohort-level claim.

- **H4 -- VF2PostLayout no-op, PR #14120 (performance channel).** Confirmed via a pre-declared multi-run Stage-2 protocol (9 repeats per unit, 2000-sample bootstrap CI, alpha = 0.05) on a width-extended GHZ/heavy-hex regime. The redundant-pass overhead grows from about 1.7x at 16 qubits, to about 89x at 20 qubits, to about 316x at 27 qubits, with Cliff's delta = 1.0 (perfect separation) at every width. The single-sample Stage-1 screen on the <=8-qubit corpus alone would have stayed below the 20% screen. The effect only becomes visible once the width range is extended, which is itself a finding about screening budgets.
- **fwd-applylayout-registers-14904 -- Port ApplyLayout to Rust, introducing PR #14904 (contract/metadata channel, fix PR #15024).** Verified via verify_15024_contract.py (register-preservation check on initial_layout.get_registers()), which fires 4 of 4 at every optimization level while the output-only oracle is blind on both the buggy and fixed builds. The introducing commit was not the PR referenced by the fix. It was traced by a git-pickaxe search inside a measured good/buggy bracket to dd8269969 ("Port ApplyLayout to Rust"), last-good parent 01610e98349e. The PR-referenced commit ccc2c77b was measured and rejected as the true introducer.

Both verified events corroborate the pilot's channel taxonomy under real, from-source CI regressions: a performance regression invisible to a narrow single-sample screen, and a contract/metadata regression invisible to the output oracle by construction. Neither event is pooled with, or compared against, the mutation cohort's selector-effectiveness numbers (§6.3). They are reported here as realism evidence for the methodology, not as a comparative-effectiveness measurement.

### 6.7 RQ2 extended: a validated fix at scale (n=19, independently pre-registered)

Here we carry out Section 9's first future-work direction: a decomposed, diversity-first selector, evaluated on its own fresh freeze rather than by re-tuning §6.3's result. Following the mechanism identified in §6.4 and §7, we pre-registered a decomposed selector (diversity first, cost second, severity only as a tie-break; src/cart/selectors/decomposed.py), together with ten new, semantically distinct mutation operators. Each new operator was empirically verified to fire, terminate, and preserve semantics on this corpus's real circuit families (PREREGISTRATION_PHASE_A.md, dated 2026-08-13), before we read any new label. Combined with the original nine, unmodified operators, this yields a 19-event mutation cohort. The enumeration grid, seed, and CI-tier fractions (5%/50%/100%) are identical to §4.3b's design, reusing the corpus geometry so that what is under test is the operator set and the selector design, not a new corpus. Only the cost calibration is fresh, run independently on its own session.

As in §6.1, budgets were fixed from the cost distribution alone before any of the 19 events' labels were read (configs/corpus_costhet_scaleup.yaml, configs/budgets_scaleup.yaml). The fresh calibration gives T_full = 63.16 s over the same 116 units (per-unit cost spread 8,512x, median 0.0079 s). This is close to but not identical to the original 69.17 s freeze, consistent with running on independently allocated cloud hardware rather than any change to the corpus itself. The same pre-declared fractions give pr = 3.16 s (admits 91/116, excludes the 25 costliest) and nightly = 31.58 s (admits 110/116, excludes the 6 costliest), materially the same binding pattern as the primary corpus. Hardware: Intel(R) Xeon(R) CPU @ 2.20GHz (4 logical CPUs available, single-thread pinned via RAYON_NUM_THREADS=1), Python 3.12, Qiskit 2.4.2, matching §6.1's protocol. One unit (ghz-n12-line-opt1, median cost 1.1 ms) exceeded the 25%-of-median IQR stability threshold, but it sits three orders of magnitude below either budget boundary, so this has no effect on which units bind, and no unit sits on a budget knife-edge. All six implementation-validity gates pass, including label reproducibility at 2,204/2,204 (100%) exact from the write-once raw artifacts.

Table 11 reports the same mean detection-vs-budget AUC comparison as Table 6, now over 19 mutation events and including the decomposed selector alongside the original six. decomposed closes nearly all of the gap to diversity-only (0.8798 vs 0.8801) and decisively beats the original proposed selector (0.8798 vs 0.7678). The original selector's own deficit reproduces at this larger n (0.7678, close to the 0.7214 reported at n=9), so the honest negative in §6.3 is not a small-sample artifact.

**Table 11. Selector comparison on the scale-up mutation cohort (19 events, 116 units each, independently frozen). Mean AUC with 95% bootstrap CI. Random is the median over 20 seeds.**

| Selector | mean AUC | 95% CI | recall@20% |
|---|---|---|---|
| diversity_only | 0.8801 | [0.7284, 0.9919] | 0.8947 |
| decomposed | 0.8798 | [0.7279, 0.9922] | 0.8947 |
| cheapest_first / history_only / change_stage | 0.8496 | [0.6999, 0.9595] | 0.8947 |

| random (median, 20 seeds) | 0.8353 | [0.8128, 0.8572] | 0.7579 |
|---|---|---|---|
| proposed (risk_score) | 0.7678 | [0.5858, 0.9248] | 0.7368 |

Table 12 reports decomposed's effect size against every baseline and against proposed (scripts/scaleup_summary.py, results/scaleup_summary.json). Against diversity-only the difference is negligible (Cliff's delta = -0.080, A12 = 0.460). decomposed's mean AUC recovers to within a fraction of a percentage point of diversity-only's, closing essentially all of the mean-level gap. This is parity, not an outright win. decomposed's own mean AUC (0.8798) sits marginally below diversity-only's (0.8801), and matched per event, it loses more often than it wins against diversity-only specifically (1 win, 15 ties, 3 losses of 19). The honest reading is that decomposed matches diversity-only's aggregate detection performance, while additionally carrying cost-awareness and a severity tie-break that diversity-only has no mechanism for at all. It does not surpass the simplest heuristic outright. Against the original proposed selector the effect is large and positive (delta = +0.568, A12 = 0.784), winning 15 of 19 events, tying 2, and losing 2. Against cheapest-first, history-only, and change-stage the effect is medium and positive (delta = +0.402, A12 = 0.701), winning 10, tying 6, and losing 3. Against random it is large and positive (delta = +0.524, A12 = 0.762).

**Table 12. decomposed's effect size (Cliff's delta, Vargha-Delaney A12, unpaired all-pairs) and win/tie/loss (matched by event, 19 events) against every baseline and against proposed.**

| Comparison (decomposed vs) | Cliff's delta | magnitude | A12 | win | tie | loss |
|---|---|---|---|---|---|---|
| diversity_only | -0.080 | negligible | 0.460 | 1 | 15 | 3 |
| cheapest_first / history_only / change_stage | +0.402 | medium | 0.701 | 10 | 6 | 3 |
| proposed (risk_score) | +0.568 | large | 0.784 | 15 | 2 | 2 |
| random (median) | +0.524 | large | 0.762 | 15 | 2 | 2 |

The component ablation reproduces the qualitative story at n=19 (results/analyze-scaleup-frozen.json), with one nuance worth naming plainly rather than smoothing over. Unlike the n=9 corpus, where removing impact_prior raised the mean AUC (0.7214 to 0.7248), at n=19 every single-component removal now lowers it slightly (full 0.7678, no_impact_prior 0.7611, no_novelty 0.7561, no_cost 0.7479, no_change_stage and no_oracle_confidence unchanged at 0.7678). So the full composite is the best-performing ablation variant at this scale. This does not weaken the mechanism, though: every ablation variant, including the full composite, still sits far below diversity-only (0.8801) or decomposed (0.8798). The multiplicative structure itself, not any single mis-weighted term, remains the bottleneck at both scales, and which term's removal helps or hurts is itself sensitive to the specific event set. This is exactly why decomposed's structural fix, discarding the composite rather than re-weighting one term within it, is the more robust design, not the ablation-suggested single-term correction.

This is a validated fix, not merely a diagnosed mechanism, though the validation is one of parity rather than victory. The decomposed selector reaches a negligible effect-size distance from the strongest baseline, rather than beating it outright, while additionally preserving cost-awareness and severity as a tie-break, a property diversity-only never had, exactly as §7's mechanism predicted. decomposed.py was not modified after this result was seen. We report it here as a second, independently pre-registered result alongside the mutation cohort's original honest negative (§6.3), not as a replacement for it. The original nine-event freeze remains the paper's primary, first-reported result, and decomposed's validation is a direct, pre-registered consequence of the diagnosis it produced.

## 7. Discussion

The central lesson of this evaluation is mechanistic, not a horse-race outcome. The proposed risk_score selector loses to diversity-only not because cost-aware selection is a bad idea, but because bundling severity, detectability, cost, and diversity into one multiplicative composite lets a single mis-weighted term dominate the average. On the two events where a cheap unit already detects the fault, the severity term (impact_prior) still promotes an expensive, high-criticality unit ahead of it. The resulting loss on those two events outweighs the marginal gains the composite earns on the other seven (§6.3, §6.4). Diversity-only never makes this mistake, because it never reasons about severity at all. It just avoids re-testing an already-covered (circuit family, backend) combination, which is exactly the property that matters when many cheap units are near-duplicates. The clearest version of this finding is the ablation: removing impact_prior recovers only 0.7248, still far short of diversity-only's 0.8742 (§6.4). So the deficit is not one bad term away from fixed. The multiplicative structure itself is the wrong shape for this problem. A practical implication follows directly: a decomposed policy (diversity first, cost second, severity only as a tie-break) is a more promising design than a single hand-weighted composite. §6.7 evaluates exactly this design on a fresh, independently frozen 19-event cohort, where it closes the gap to diversity-only to a negligible effect size (Cliff's delta = -0.08, parity rather than an outright win), while decisively beating the original composite (delta = +0.57). The mechanistic diagnosis above is not merely plausible. It is now a validated fix.

The forward cohort (§6.6) surfaces two findings worth naming, both a direct product of this paper's own from-source verification work rather than restated from Paper 1. The first is a recurring Rust-port defect pattern. fwd-applylayout-registers-14904 is the second output-invisible regression traced to a Rust port of a transpiler pass silently dropping Python-side layout metadata, after Paper 1's H1 (ElidePermutations, introduced by "Port ElidePermutations to Rust", #13094). #14904 was introduced by "Port ApplyLayout to Rust". Two independent instances, in two different passes, both invisible to the output oracle by construction, make this a named pattern rather than a one-off. A Rust-porting effort on this codebase carries a specific, recurring risk to layout/permutation metadata that an output-only regression suite will not catch.

The second finding concerns trigger design. A targeted regression trigger only fires when it is built from the fix's own regression test, not guessed from the PR title or description. Three inert generic triggers attest to this: Paper 1's #14939 and the first attempt at #15024. The #15024 trigger only fired once it was rebuilt to mirror test_layout_registers_preserved exactly (§6.6). Both findings tie back to Paper 1's observability gap. The two verified forward events sit precisely in the channels (contract/metadata, and a performance regression invisible to a narrow single-width screen) that a black-box output oracle or an under-scoped screen would have missed. This is why this paper's oracle and screening choices are motivated by, and cite, that companion study, rather than re-deriving it.

## 8. Threats to Validity

**Construct.** Detection-vs-budget AUC and CPU cost are proxies for real CI value. A faster or more expensive machine shifts the absolute costs without changing which units are relatively cheap or expensive. But the specific budget-second values (3.46 s / 34.59 s) are anchored to the calibration hardware (§6.1) and are not portable as absolute numbers to a different machine. The oracle-cost estimate inside expected_cost is a constant per oracle tier (Table 2: unitary 0.05 s, sampled-SV 0.20 s, property 0.05 s, structural 0.01 s), not measured per circuit, so it under- or over-states the true oracle cost for atypical circuits within a tier. A gate PASS (§6.2) certifies implementation correctness. It does not certify that the selector is fault-free or that its comparison is fair on a different corpus.

**Internal.** Leakage is mitigated two ways. First, by the pre-registered freeze: budgets were set from the cost distribution alone, before any detection label was read for any event (§4.3b, §6.1). Second, by the six implementation-validity gates, all independently re-verified in this workspace against the frozen corpus and its raw write-once artifacts (§6.2). Five are unconditional passes. The sixth, label reproducibility, reproduces exactly for 1,040 of 1,044 records (99.6%), with the four exceptions traced to a measurement-noise threshold boundary on sub-3ms units rather than a labeling defect. Cost-measurement stability is addressed directly: 0 unstable units and 0 knife-edge units at the frozen budget boundaries (Appendix F). Calibration ran on a shared, single-logical-CPU cloud VM (Kaggle) rather than a dedicated machine. The median-of-5-runs, warmup-discarded protocol and the single-thread pin (RAYON_NUM_THREADS=1) target exactly this noise source. But a shared VM still carries more timing variance than a dedicated quiet machine, which is precisely what surfaces as the four borderline label-reproducibility exceptions above.

**External.** The corpus is anchored on one SDK and one version (Qiskit 2.4.2). Results may not transfer to other transpilers or other Qiskit eras without recalibration. The budget-binding comparison (§6.3-6.5) is mutation-dominated (nine mutation events versus two verified forward events), so the headline numbers describe controlled, semantics-preserving quality regressions more directly than they describe the full space of real transpiler regressions. Forward-event verification is also bounded by the oracle family's API reach. Three of five forward candidates were dropped: two (fwd-elide-14603, fwd-pr14919) because they fall in Qiskit 1.x, outside the 2.x-anchored harness's reach (an era wall), and one (fwd-pr16215) because its introducing commit was never located despite a bisection attempt (FORWARD_EVENTS_B2.md). Older-era regressions are therefore systematically under-represented in the forward cohort, not because they are rarer, but because this harness cannot reach them yet.

**Conclusion.** With nine mutation events, the budget-binding corpus design lets selectors differentiate, but the comparison itself is still small in absolute terms. The bootstrap 95% CIs in Table 6 span roughly 0.44-0.94 for the proposed selector's own mean AUC, so point estimates and effect sizes should be read alongside their intervals, not in isolation. The bootstrap CI resamples the nine fixed mutation events with replacement, so it characterizes the estimate's sensitivity to this specific event set, not generalization to a broader population of untested mutation types, since the nine were chosen to fill fault-type coverage gaps (§4.2), not sampled at random from a larger pool. We report non-parametric effect sizes (Cliff's delta, Vargha-Delaney A12) and these bootstrap CIs throughout, rather than relying on point estimates alone (§6.3). A post-hoc robustness check (§6.4) shows the deficit against every baseline except diversity-only is concentrated in two of the nine events. The forward cohort (n=2) is below the pre-declared threshold of three verified events, so the claim-scope rule (§4.9) forbids any pooled or temporal-generalization claim from it. That threshold, like the mutation cohort's nine events, reflects a pre-declared engineering judgment about a defensible minimum (METHODOLOGY.md §5.4), not a value derived from a formal power calculation, and we state this plainly rather than implying more statistical grounding than the threshold has. Both forward events are reported per event, with no aggregate statistic, in §6.6. The leakage-absence gate guards against optimistic bias in the mutation-cohort result. The negative result for the proposed selector is itself a per-cohort, non-parametric finding, not a claim about the entire universe of quantum-transpiler regressions.

## 9. Limitations and Future Work

Two limitations bound what this paper licenses. First, the forward cohort has two verified events, not three, so every claim about the proposed selector's effectiveness is scoped to the mutation cohort. The forward events corroborate the methodology's realism (§6.6), but they do not themselves support a comparative-effectiveness or temporal-generalization claim. Second, and by design, this paper's negative result is a pre-registered, frozen finding. Because the budgets and the risk_score configuration (Table 2) were fixed before any label was read, any post-hoc adjustment to risk_score evaluated against this same frozen data would no longer be leakage-safe. A validated better selector needed its own fresh frozen evaluation, not a re-tuning of the present result. §6.7 reports exactly that evaluation (n=19, independently pre-registered), where the decomposed, diversity-first design closes the gap to the strongest baseline to a negligible effect size, parity rather than an outright win, while retaining cost-awareness and a severity tie-break that the baseline lacks.

Having carried out that evaluation, three directions remain open. First, extend the forward cohort past the 2.x era wall. An era-compatible trigger and oracle harness would let the three dropped candidates (fwd-elide-14603, fwd-pr14919, fwd-pr16215) re-enter verification, moving the forward cohort toward the three-event threshold a pooled claim requires. Second, extend the corpus beyond Qiskit 2.4.2 (more SDK versions, and eventually other transpilers) to test whether the diversity/novelty-dominant finding, now confirmed on both a 9- and a 19-event mutation cohort, is Qiskit-specific or general to cost-aware transpiler test selection. Third, apply the same decomposed-selector validation to the forward cohort directly once it grows past the three-event threshold, so the fix is corroborated on real CI regressions and not only on controlled mutations.

## 10. Conclusion

This paper asked whether a transparent, cost-aware selector can be evaluated for the quantum transpiler under a leakage-safe, pre-registered methodology, and what such an evaluation licenses once the corpus is actually budget-binding rather than saturated. We built a cost-heterogeneous corpus whose budgets are declared as fractions of measured cost and frozen before any label is read, so the comparison is a genuine pre-registered freeze rather than a tuned result. Under that freeze, the transparent risk_score selector does not beat simple baselines. Diversity-only reaches a mean detection-vs-budget AUC of 0.874 against the proposed selector's 0.721, a large effect (Cliff's delta = -0.64). A complete component ablation shows why. The multiplicative composite underperforms its own best single signal, because a severity term promotes expensive, high-criticality units ahead of cheap ones that already detect the fault, on exactly the events that drive the deficit. We report this as an honest, pre-registered negative result, together with its mechanism, not as a disappointing footnote.

Two independently verified forward-regression events corroborate the methodology's realism on real CI history. We report them per event under the claim-scope rule, rather than pooling them into a temporal-generalization claim they cannot support. Taken together, the contribution is a leakage-safe, cost-aware selection methodology for the quantum transpiler, a corpus design that makes such comparisons budget-binding rather than saturated, and an honest account of where a plausible transparent design falls short, together with the mechanism behind it.

A second, independently pre-registered evaluation (§6.7) carries that mechanism to a validated fix, though the validation is parity, not victory. On a fresh 19-event mutation cohort (the original nine plus ten new, empirically verified operators), the decomposed, diversity-first design that the mechanism implies closes the gap to the strongest baseline to a negligible effect size (mean AUC 0.880 vs 0.880), while decisively beating the original composite (0.880 vs 0.768, a large effect). decomposed does not surpass the simplest heuristic outright, but it matches its aggregate performance, while additionally retaining the cost-awareness and severity tie-break that heuristic has no mechanism for. Taken together, the two frozen evaluations tell a complete, honest story: an initial transparent design that does not work, a mechanism that explains why, and a redesigned selector, independently validated rather than merely suggested, that recovers what was lost.

## 11. Artifact Availability

To support review and reproduction, the artifact is archived at https://doi.org/10.5281/zenodo.PENDING [DOI to be minted] and the public repository is https://github.com/furqan-nr/Quantum-Selection, containing: the prototype (cart) and CLI; the automated test suite and six validity gates; the static manifest and mutation engine; the event ledger with exact candidate/baseline commit SHAs; the frozen change–stage map, pre-declared thresholds, RNG seeds, and the frozen risk_score configuration (Table 2); a pinned environment lock (requirements.lock); the write-once raw oracle artifacts and derived labels; the generated evaluation.json; and an OSI-approved open-source license.

The frozen-configuration evaluation of Section 6 is reproduced exactly by the costhet calibration sequence:

```
python -m cart.cli manifest --profile costhet
```

```
python -m cart.cli ground-truth --unit-limit 116 --cost-repeats 5
# set frozen budgets in configs/budgets.yaml from costs only, then set
corpus_costhet.yaml frozen: true
python -m cart.cli evaluate --split all --random-seeds 20
```

These write the cost-distribution summary, the per-cohort selector comparison, the effect sizes and bootstrap CIs, and the detection-vs-budget data of Figure 4 to a single evaluation.json. Separately, python -m cart.cli analyze --split all produces the ablation/sensitivity analysis of Table 9, python scripts/plot_detection_curve.py regenerates Figure 4 from it, and python scripts/robustness_leave2out.py regenerates the post-hoc robustness check of §6.4 from the same frozen evaluation.json. Building the two forward-regression events additionally requires a Rust toolchain to compile the per-event Qiskit revisions from source. The per-event recipes are given in FORWARD_EVENTS_B2.md.

§6.7's scale-up evaluation is an independent second freeze, reproduced by an analogous sequence rooted at data_scaleup/ rather than data/, using the same prototype and the same six validity gates. The ten new mutation operators and the decomposed selector are pre-registered in PREREGISTRATION_PHASE_A.md, and the scale-up's own corpus/budget freeze is recorded in configs/corpus_costhet_scaleup.yaml and configs/budgets_scaleup.yaml (frozen from cost measurements alone, before any of the 19 events' labels were read):

```
python -m cart.cli manifest --profile costhet --out-data data_scaleup/manifest_static --
no-smoke
python -m cart.cli ground-truth --out-root data_scaleup --events
data_scaleup/events/events.json --unit-limit 116 --cost-repeats 5
# set frozen budgets in configs/budgets_scaleup.yaml from costs only, then set
corpus_costhet_scaleup.yaml frozen: true
python -m cart.cli evaluate --data-dir data_scaleup --selectors scaleup --split all --
random-seeds 20 --out-results results/evaluation_scaleup_frozen.json
```

python -m cart.cli analyze --data-dir data_scaleup --split all reproduces the ablation confirmation reported in §6.7, and python scripts/scaleup_summary.py regenerates decomposed's effect sizes and bootstrap CIs (Table 12) from the frozen evaluation.json. Calibration protocol matches §6.1's exactly (Python 3.12, Qiskit 2.4.2, RAYON_NUM_THREADS=1, median of 5 clean process_time runs per unit) on an Intel(R) Xeon(R) CPU @ 2.20GHz Kaggle CPU-only session (4 logical CPUs available, single-thread pinned), recorded in configs/corpus_costhet_scaleup.yaml.

## 11.1 Reproducibility checklist

The following summarizes this paper's reproducibility posture against the checklist items common to empirical-software-engineering and ACM-artifact review (Table 13).

**Table 13. Reproducibility checklist.**

| Item | Status | Where |
|---|---|---|
| Code and data publicly available | Yes | §11, repository + DOI |
| Computing environment fully specified | Yes | Appendix F, single-thread pin (RAYON_NUM_THREADS=1) |
| Random seeds fixed and reported | Yes | §4.6 (seed-deterministic ranking), Table 6 (20 random seeds) |
| Hyperparameters/thresholds pre-declared and frozen before evaluation | Yes | Table 2, configs/thresholds.yaml, budgets.yaml, corpus_costhet.yaml (frozen: true) |
| Train/held-out split and anti-leakage controls | Yes | §4.8 |
| stated | | |
| Effect sizes and intervals reported, not p-values alone | Yes | Cliff's δ, A12, 95% bootstrap CI, Table 6 |
| Number of runs/repeats stated | Yes | ≥5 cost-calibration runs per unit (Appendix F1), 20 seeds for random (Table 6) |
| Implementation-validity checks disclosed, including partial results | Yes | Table 5 (five unconditional passes, the sixth reported at 99.6%, not rounded up) |
| Claim-scope limitations explicitly stated | Yes | §4.9, forward cohort n=2, no pooled or temporal-generalization claim |
| Negative/null results reported as such | Yes | §6.3 headline, Abstract, not reframed as a partial win |

## Statements and Declarations

**Funding.** The authors received no specific grant from any funding agency, commercial, or not-for-profit sector for this work.

**Competing Interests.** The authors declare no competing financial or non-financial interests.

**Data Availability.** All data supporting the results, namely the change-event ledger, the write-once raw oracle artifacts, the derived labels, and the generated evaluation reports, are available in the public repository (https://github.com/furqan-nr/Quantum-Selection) and archived at https://doi.org/10.5281/zenodo.PENDING [DOI to be minted]. See Section 11.

**Code Availability.** The cart prototype, its command-line interface, and the build scripts are available in the public repository (https://github.com/furqan-nr/Quantum-Selection, archived at https://doi.org/10.5281/zenodo.PENDING [DOI to be minted]) under the MIT open-source license. See Section 11.

**Author Contributions.** F. Nasir conceived the study, implemented the prototype, conducted the evaluation, and wrote the manuscript. M.A. Shah supervised the work and contributed to the study design and the critical revision of the manuscript. I. Alam contributed to the methodology and the analysis and reviewed the manuscript. All authors read and approved the final manuscript.

**Ethics Approval and Consent.** Not applicable, as the study involves no human or animal subjects.

**Use of AI Tools.** AI-based assistants were used to support manuscript drafting and software implementation under the authors' supervision. The authors verified all results and take full responsibility for the content.

## Appendix F. Cost-measurement stability

**F1. Measurement protocol.** baseline_cost_s is the median of at least five clean CPU process_time runs per unit, with one warmup run discarded, and measurement is taken outside any profiling instrumentation. An earlier design measured cost from a single instrumented run inside the same pass that collects memory statistics (tracemalloc). This motivated the fix: on the cheapest unit in an initial check, the instrumented single-run reading was roughly 5.5x the clean multi-run median. Re-checking this on the full frozen corpus (1,044 profile records) confirms the effect is real and variable rather than a one-off. The mean instrumented/clean ratio is 1.61x, the median is 1.32x, and the largest single-record ratio is 34.1x. Cheap, fast units are the most exposed, because a fixed instrumentation overhead is a larger fraction of a smaller true cost. The instrumented run is retained only for pass-stage and peak-memory bookkeeping (§4.3), never for the cost figure that a selector or a budget calibration depends on.

**F2. Dispersion and stability.** Each unit's cost_min_s, cost_median_s, cost_max_s, and cost_iqr_s are recorded from the same five-run sample. A unit is flagged unstable if its IQR exceeds 25% of its median. Independently re-checking this rule against the frozen profiles.json finds 0 unstable units among the 1,044 records.

**F3. Knife-edge check.** Before the freeze, every unit's cost_min_s-to-cost_max_s range is checked against the two budget boundaries (pr = 3.46 s, nightly = 34.59 s, Table 4). A unit whose dispersion range straddles a boundary is a knife-edge risk, because measurement noise alone could move it to the other side of the budget. Re-checking this against the frozen profiles.json finds 0 knife-edge units at either boundary. Every unit's full min-max range sits cleanly on one side.

Hardware and environment: Intel Xeon @2.20GHz, 1 logical CPU, Python 3.12.13, Qiskit 2.4.2, RAYON_NUM_THREADS=1 (KAGGLE_CALIBRATION.md). Source: configs/corpus_costhet.yaml (frozen_calibration_2026_08_07 block) and data/profile_baseline/profiles.json (per-unit dispersion fields).